
\input phyzzx
\input macros
\overfullrule=0pt

\rightline{UCLA/92/TEP/31}
\rightline{UFIFT-HEP-92-33}
\bigskip

\centerline{{\bf CASIMIR FORCES BETWEEN BEADS ON STRINGS }
\footnote{*}{Research Supported in part by NSF grant PHY-89-15286
and by DOE contract No. DE-FG05-86ER40272.}}
\bigskip

\centerline{{\bf Eric D'Hoker}
\footnote{**} {Electronic Mail Address: DHOKER@UCLAHEP.BITNET.}}
\centerline{{\it Physics Department}}

\centerline{{\it University of California, Los Angeles}}
\centerline{{\it Los Angeles, California 90024-1547, USA}}

\bigskip
\centerline{{\bf Pierre Sikivie}
\footnote{***} {Electronic Mail Address: SIKIVIE@UFHEPA.PHYS.UFL.EDU}}
\centerline{{\it Physics Department}}
\centerline{{\it University of Florida}}
\centerline{{\it Gainesville, Florida 32611, USA}}

\bigskip

\centerline{\bf ABSTRACT}

We consider a string with uniform energy density and tension and with a
number of pointlike masses attached at fixed interdistances.  We
evaluate the effective interaction forces between these masses
induced by the quantum fluctuations of the string.
For interdistances large compared to the thickness of the
string and small compared to the total length,
these forces are universal and attractive and fall off, for large distances
as $1/r^3$ and for small distances as $1/r$.
The attractive nature of these forces creates an instability
under which masses added to the string tend to aggregate.

\vfill\eject

In most dynamical models of string evolution, it is assumed from the
outset that the energy density and tension are constant along the
string.  It is within this framework that the Nambu-Goto [1] and Polyakov [2]
models are formulated, which are the basis of the string theory of
elementary particles and their interactions.  Also the extension to
include elastic interaction energy  that opposes bending of the string [3]
and the models considered in [4] make this assumption.

There may be applications of string
dynamics where this assumption is too stringent. The
energy density and tension generally varies along strings which have zero
modes attached to them, or strings which are wiggly on scales much
smaller than the scale of interest to the observer.  In
particular, such may be the
case with QCD strings, cosmic gauge strings, and the vortices that occur
in superconductors and superfluids.  Also,
in applications to polymer
physics, the mass distribution along a string is generally uneven.
A natural extension of the Nambu-Goto model was
proposed at the classical level in which new degrees of freedom on the
string allow for variable energy density and string tension [5].

In the present paper, we examine the forces induced by
the transverse quantum mechanical fluctuations of
the string on
deviations away from constant energy density.
Inhomogeneities in the energy density of the
string alter the frequency spectrum of the transverse string
oscillations.
This results in
Casimir type forces between the inhomogeneities.
In this letter, we evaluate these forces in the
case of pointlike masses attached to a straight string.

Part of our motivation is a possible connection with
gravity.  We will find below that when two
inertial masses, say $m_1$ and $m_2$, are attached to the string,
quantum fluctuations are responsible for an attractive force
between the masses which is
proportional to the product $m_1 m_2$ in the limit of small masses.
The force does not scale like $1/r^2$ as in gravity, but it is
proportional to
the product of inertial masses, so that an analogue
of the equivalence principle holds here, which
arises naturally from quantum mechanical effects.

Our assumptions are as follows.  We consider a straight
string of length $L$ and thickness $1/\Lambda$,
with constant energy density $\rho$ and constant tension $\kappa$.
$N$ masses $m_i$ are attached to the string in various
fixed locations with coordinates $x_i~(i=1, \cdots , N)$.
Distances between the various masses will always be taken to be
much smaller than the total length $L$ [6], but much larger than
the thickness $1/\Lambda$.
In this limit, the Casimir forces will be found to be independent
of both $L$ and $\Lambda$.
Only the $d-1$ transverse degrees of oscillation of the non-relativistic
string in $d$-space dimensions are retained.
We include a tension term in the action, but we shall neglect
effects due to the bending of the string.
The dynamics are thus assumed to be governed by
$$
S = \int dt \biggl \{ \int\nolimits_0^L dx  \half (\rho
{\dot\varphi}^2 - \kappa\varphi '^2 ) + \half \sum_{i=1}^N m_i
{\dot\varphi}^2(x_i) \biggr \} \eqno (1)
$$
where $\varphi$ stands for the $d-1$ transverse oscillation degrees of freedom.

We will encounter some
quantities which require regularization.  This may be effected by a
cutoff on the wave vectors of the transverse string oscillations.
We may in some applications think of this cutoff as physical in nature.
At momenta beyond the cutoff, one would be probing the internal structure of
the string, which typically involves new physics, not summarized by Eq.(1).
Thus the ultraviolet cutoff should be of the order of the inverse
thickness of the string, and we take its value to be $\Lambda$.

First, we derive a general equation for the frequencies of the
oscillation eigenmodes.  Let us label masses $m_i$ such that
their locations on the string are ordered:
$0=x_o  < x_1 < x_2 < x_3 < \cdots <x_N <  x_{N+1}=L$.
The equation for a mode with frequency $\omega$ is derived from (1):
$$
\kappa \varphi_\omega^{''}(x) + \rho \omega^2\varphi_\omega(x)
= - \omega^2
\sum_{i=1}^N m_i \varphi_\omega(x_i)\delta(x-x_i)
\eqno(2)
$$
Within each interval between the masses, $\varphi$ is just a sum of two
exponentials:
$$
x_{j-1} \leq x \leq x_j \qquad \varphi_\omega(x) = A_j(k) e^{ikx} +
B_j(k) e^{-ikx}
\eqno (3)
$$
where $\rho\omega^2 = \kappa k^2$.
Continuity of $\varphi_\omega$ and the
known discontinuity of $\varphi'_\omega$ lead to the following
transfer equations ($i=1, \cdots , N$):
$$
\pmatrix{A_{i+1}\cr
         B_{i+1}\cr}  = T_i \pmatrix{A_i\cr B_i \cr} \qquad \qquad
\pmatrix{A_1\cr B_1 \cr}  =  T_0 \pmatrix{A_{N+1}\cr
        B_{N+1}\cr}\ .
\eqno(4)
$$
Here we define the transmission matrices:
$$
T_j  \equiv \pmatrix{ 1 + t_j & t_j~e^{-2ikx_j}\cr
                -t_j~e^{+2ikx_j} & 1-t_j\cr} \qquad \qquad
T_0  \equiv \pmatrix{ e^{ikL}& 0\cr 0 & e^{-ikL}\cr}
\eqno(5)
$$
with $t_j = {ik m_j \over 2 \rho}$.
The equation for the momenta $k$ is the condition that the system of
Eqs. (4)
admits non-zero solutions :
$$
\det (1-T_N T_{N-1} \cdots T_1 T_0) = 0
\eqno (6)
$$
We want to solve Eq.(6) and sum the
corresponding oscillation frequencies.
We will encounter infinities (i.e. quantities which are
infinite in the limit $\Lambda \to \infty$) but they
can all be absorbed into renormalizations of the parameters that have
been introduced so far :  energy density $\rho$, tension $\kappa$,
and the masses $m_i$ of the extra particles on the string.
After these renormalizations have been effected, we will
evaluate the interaction forces between
the masses on the string.

When no masses are present $(N=0)$, the momenta are given by $k_n^\pm L =
2\pi n$ for $n = 0, 1, 2, \cdots$.  There are two linearly
independent modes with the same frequency for every
positive $n$.
When masses are added to the string, this degeneracy is lifted as
will be described below.  The zero point energy that results
from summing the eigenfrequencies is proportional $L$ and
amounts to a renormalization of the effective energy per unit length
$\rho$:
$$
\delta\rho = + 2\pi \hbar
\biggl ({\kappa\over\rho}\biggr )^\half \Lambda^2 (d-1)
\eqno(7)
$$
When a single mass $m$ is present $(N=1)$, Eq. (6) becomes:
$$
0 = 2-2\cos kL + {m\over \rho} k \ \sin kL
\eqno (8)
$$
The problem of summing the solutions of this equation
simplifies when the limit $L\to \infty$ is taken.  We define phase shifts as
follows:
$$
k_n^\pm L = 2 \pi n + 2\delta_n^\pm~,~~~~ \qquad n =
0,1,2,\cdots~,
\eqno (9)
$$
and solve Eq. (8) in terms of $\delta_n^\pm$ as $L \to \infty$ for fixed $k$:
$$
\cases{\delta_n^+ = 0\cr
        \delta_n^- = - {\rm Arctg}~~ \bigl ( {m\over \rho L}\pi
n\bigr)\cr}\ .
\eqno(10)
$$
Note that the unshifted momenta $k_n^+$ correspond to modes of oscillation
which have a node at the location of the mass $m$ on the string.
The energy shift which results from adding the mass $m$
is independent of the length $L$ (for large $L$), and corresponds to a
renormalization of the effective mass due to string fluctuations:
$$
\delta m = - {\hbar\over 2\pi}
\biggl({\kappa\over\rho}\biggr)^\half \left[ \Lambda {\rm Arctg}
\biggl({\Lambda m\over 2\rho}\biggr) - {\rho\over m} \ell n
\left(1 + \biggl({\Lambda m\over 2\rho}\biggr)^2\right)\right]
(d-1)\ .
\eqno (11)
$$

When two masses are present $(N=2)$, Eq. (6) becomes:
$$
0=2-2\cos kL + {m_1+m_2\over \rho} k \sin kL - {m_1m_2\over
\rho^2} k^2\sin kx \sin k(L-x)\ ,
\eqno(12)
$$
where $x = x_2 - x_1$.  We parametrize the solutions again as in Eq. (9).
One can show that the
phase shifts $\delta_n^\pm$ are bounded in magnitude by $3\pi/2$
when $m_1$ and $m_2$ are varied from zero to infinity.  Hence, in the
limit $L \to \infty$ with $k$ fixed, we can neglect $\delta_n$
compared with $\pi n$ wherever $k$ appears
in Eq. (12) other than in the combination $\sin kL$ or $\cos kL$.
Eq. (12) is then quadratic in ${\rm tg}  \delta_n$.  It can be solved
explicitly, and we find:
$$
\delta_n^+ + \delta_n^- = - {\rm Arctg}
{\pi n (m_1+ m_2)\rho L - \pi^2 n^2 m_1m_2 \sin 4\pi n x/ L
\over \rho^2L^2-\pi^2n^2 m_1m_2 ( 1-\cos 4 \pi n x/ L) }\ .
\eqno (13)
$$
{}From
the requirement that $\delta_n^+ + \delta_n^-$ be a continuous
function of $k_1, m_1,~m_2$ and $x$ it follows that
the Arctg function takes values in the interval $\biggl[ -
{\pi\over 2},~{\pi\over 2}\biggr]$ when the denominator $\rho^2
L^2 - \pi^2 n^2 m_1 m_2 \biggl(1-{\rm cos} {4\pi n x\over L}\biggr)$ is
positive, and in the interval $\biggl[{\pi\over 2},~{3\pi\over
2}\biggr]$ when that denominator is negative.

Thus, the shift in zero point energy when any two masses are
added to the string is given in the
$L\to\infty$ limit by
$$
\delta E = - (d-1){\hbar\over 2 \pi} \biggl ({\kappa\over \rho}\biggr )^\half
\int\nolimits_0^\infty dk ~f\biggl({k\over\Lambda}\biggr) {\rm Arctg}
{2 k(m_1+ m_2)\rho - k^2m_1m_2 \sin 2 k x \over
4\rho^2-k^2m_1m_2 (1-\cos 2k x)}\ ,
\eqno(14)
$$
where the $f(x)$ is a smooth function approximating the step
function $\theta (1-x)$.
Actually, this energy shift depends on the cutoff $\Lambda$
only through $x$-independent terms.
It is easy to show this order by order in an expansion
in small masses, and one can show this also more
generally for all values of mass.
Hence one of the central results
of this paper : the interaction forces are universal,
i.e. independent of $L$ and $\Lambda$ for large L and $\Lambda$.
On dimensional grounds, we have then:
$$
\delta E = \delta E_0 + V(x) \qquad\qquad
V(x)  = {1 \over x} F({m_1m_2\over \rho ^2 x^2}, {m_2 \over
m_1})\ .
\eqno (15)
$$
There is a set of interesting limits that can be taken in this
expression.

The effect of small masses $m_1, m_2 \ll \rho x$
is easily evaluated by expanding Eq.(14) in a power series.
We find, in this regime, a universal two body
interaction potential, proportional to the two masses $m_1$ and $m_2$
and falling of as the inverse cube of distance:
$$
V(x) = - (d-1){\hbar\over 32 \pi} {m_1m_2\over \rho^2} \biggl
({\kappa\over \rho}\biggr )^\half \bigl [ {1\over x^3} -{3\over 4}
{m_1+m_2\over \rho}~ {1\over x ^4} \bigr ] +{\cal O} ({m^2 \over
x^5})\ .
\eqno(16)
$$
The effect of one small mass $m$ to first order in that mass
but all ordes in the other mass $M$ is also interesting and given
by the following expression
$$
V(x) = {\hbar \over 2 \pi} \biggl ( {\kappa \over \rho}
\biggr ) ^\half (d-1) m \int_0^\infty dk {k^2M\over 4\rho^2+ k^2
M^2} \biggl ({kM\over 2\rho} {\rm cos} 2kx + {\rm sin} 2kx\biggr)\ .
\eqno (17)
$$
In the limit of small $M$, we recover Eq.(16), and in the limit of infinite
$M$, we obtain
$$
V(x) = -(d-1){\hbar m \over 16 \pi \rho} \biggl ( {\kappa \over \rho }
\biggr )^\half {1 \over x^2}\ .
\eqno (18)
$$
In the limit where both masses are large, we have
$$
V(x) = - (d-1)\hbar \biggl ({\kappa\over \rho}\biggr )^\half
\bigl [ {\pi \over 24x} - {\rho (m_1 + m_2)\over 2\pi m_1 m_2}
\ell n~x + {\cal O} \biggl({x\rho^2 \over m^2}\biggr)
\bigr ] \ .
\eqno (19)
$$
The $1/x$ leading part of this potential was derived
originally in [7], where it was also shown to be universal.
Eq. (14) is a systematic expression for corrections
to this leading behavior.

The equation for a general number $N$ of additional masses can also be
readily derived from Eq. (6), and reads (with $x_{ij}=x_j-x_i$) :
$$
0 = 2-2\cos kL - \sum_{p=1}^N \biggl({-k\over \rho}\biggr)^p
\sum_{(\alpha_1\cdots\alpha_p)} \biggl( \prod_{i=1}^p m_{\alpha_i}\biggr)
\biggl(\prod_{i=2}^p \sin kx_{\alpha_{i-1}\alpha_i}\biggr) \sin k
\biggl(L-x_{\alpha_1 \alpha_p}\biggr)
\eqno (20)
$$
where the $(\alpha_1 \cdots\alpha_p)$ are all possible ordered
$(\alpha_1 < \alpha_2 \cdots < \alpha_p)$ subsets
with $p$ elements of (1,2...N).
This equation can also be solved in the limit where $L\to
\infty$, and one finds:
$$
\delta_n = \delta_n^+ + \delta_n^- = {\rm Arctg} ~~{a_{\cos} \over
2- a_{\sin}}
\eqno (21)
$$
where
$$
a_{\left (\cos\atop
        \sin\right )}
= \sum_{p=1}^N~\biggl({-k\over \rho}\biggr)^p~~
\sum_{(\alpha_1\cdots\alpha_p)}
\biggl(\prod_{i=1}^p m_{\alpha_i}\biggr) \biggl(\prod_{i=2}^p
\sin kx_{\alpha_{i-1} \alpha_i}\biggr) ({\cos \atop \sin } )
(kx_{\alpha_1\alpha_p})
\eqno(22)
$$
and the energy shift may be written down in analogy with Eq. (14).
The three body forces may be of special interest.  They are given
in the limit of small masses by
$$
V(x_{12}, x_{13},x_{23})
= +(d-1) {3\over 64 \pi} \hbar \biggl ({\kappa\over \rho}\biggr
)^\half {m_1m_2m_3\over \rho^3}
{1\over x_{13}^4}
\eqno(23)
$$
for three masses with ordering $x_1 < x_2 < x_3$.  Note that
two body forces
dominate over three-body and higher forces at large distances.
One can also evaluate the three body forces in the
large mass (i.e. small distance) limit.
It is found that the two-body forces dominate in
that limit as well.

We conclude with some additional remarks
about the nature of these Casimir type forces.
The most interesting
physical implication of our results is perhaps that all forces being
attractive between all beads, an instability occurs
along the lines of the instability present
in gravity.
Masses attached to a straight string tend to aggregate.
The forces we have investigated are similar to gravity in other
respects as well.  Indeed we found that the force between two
inertial masses $m_1$ and $m_2$, attached to a straight string,
is proportional to the product $m_1 m_2$ in the limit of small
masses.
Therefore this force
obeys the equivalence principle in that limit.
One may even hope to
explain the phenomenon of gravity by postulating that we and
everything we know of are attached as zero (or near zero) modes
to a flat 3+1 dimensional hypersheet embedded in a higher dimensional space.
However, we found that the force between the two point-like
masses on the string, in the limit of small masses where the
``principle of equivalence" holds, is proportional to
$ - m_1 m_2 \over \rho ^2  r^4 $, where $r$ is the
distance between the masses. Its naive generalization
to 3+1 dimensions,  $ - m_1 m_2  \over  \epsilon ^2 r^8 $, where
$\epsilon$ is the energy per unit volume of the hypersheet, does not
have the desired $r$-dependence.
It is tantalizing however that the equivalence
of inertial and ``gravitational" masses in the toy model
arises naturally from quantum mechanical effects.
\bigskip

\noindent {\bf Acknowledgements:}

We would like to thank the Institute for Theoretical Physics at
UC Santa Barbara and the Aspen Center for Physics
for providing the forum where this work was
initiated and for their hospitality.
\bigskip

\noindent {\bf References:}

\item{[1]} Y. Nambu, in {\it Symmetries and Quark Models}
Proceedings of International Conference, Wayne State University,
1969, edited by Ramesh Chand (Gordon and Breach, New York) 1970;
T. Goto, Prog. Theor. Phys. 46 (1971) 1560
\item{[2]} A.M. Polyakov, Phys. Lett. B103 (1981) 207
\item{[3]} A.M. Polyakov, Nucl. Phys. B268 (1986) 406
\item{[4]} J. Polchinski and A. Strominger, Phys. Rev. Lett. 67 (1991) 1681
\item{[5]} J. Hong, J. Kim and P. Sikivie, Phys. Rev. Lett. 69
(1992) 2611
\item{[6]} Open strings with masses attached at their ends were
considered as elementary particle models in: A. Chodos and C. B.
Thorn, Nucl. Phys. B72 (1974) 509; I. Bars and A.J. Hanson, Phys. Rev. D13
(1976
)
1744; W.A. Bardeen, I. Bars, A.J. Hanson and R.D. Peccei,
Phys. Rev. D13 (1976) 2364
\item{[7]} M. L\"uscher, Nucl. Phys. B180 (1981) 317

\end

\vfill\eject
\end